**Definitive Evidence of Interlayer Coupling Between Ga$_{1-x}$Mn$_x$As Layers Separated by a Non-Magnetic Spacer**

B.J. Kirby & J.A. Borchers, NIST Center for Neutron Research, National Institute of Standards and Technology, Gaithersburg, MD 20899

X. Liu, Z. Ge, Y.J. Cho, M. Dobrowolska & J.K. Furdyna, Department of Physics, University of Notre Dame, Notre Dame, IN 46556

**Abstract**

We have used polarized neutron reflectometry to study the structural and magnetic properties of the individual layers in a series of $Al_yBe_zGa_{1-y-z}As/Ga_{1-x}Mn_xAs/GaAs/Ga_{1-x}Mn_xAs$ multilayer samples. Structurally, we observe that the samples are virtually identical except for the GaAs spacer thickness (which varies from 3-12 nm), and confirm that the spacers contain little or no Mn. Magnetically, we observe that for the sample with the thickest spacer layer, modulation doping by the $Al_yBe_zGa_{1-y-z}As$ results in $Ga_{1-x}Mn_xAs$ layers with very different temperature dependent magnetizations. However, as the spacer layer thickness is reduced, the temperature dependent magnetizations of the top and bottom $Ga_{1-x}Mn_xAs$ layers become progressively more similar – a trend we find to be independent of the crystallographic direction along which spins are magnetized. These results definitively show that $Ga_{1-x}Mn_xAs$ layers can couple across a non-magnetic spacer, and that such coupling depends on spacer thickness.

PACS: 75.50.Pp, 73.61.Ey, 61.12.Ha

Interlayer coupling in magnetic multilayer structures is a subject of important basic and applied research interest.[1] The most famous example of this is the giant magnetoresistance effect[2], which is due to interlayer exchange coupling between ferromagnetic (FM) metal layers across a non-magnetic spacer layer, and has been exploited in a multitude of devices.[3] With the recent advent of artificial dilute FM semiconductors such as $Ga_{1-x}Mn_xAs$,[4,5,6] attention has been given to interlayer coupling in all semiconducting multilayer structures.[7] SQUID magnetometry,[8,9,10,11] magneto-transport measurements,[8,9,11] and qualitative analysis of neutron diffraction superlattice peaks[12,13] have been used to produce some evidence of interlayer coupling in $Ga_{1-x}Mn_xAs$ based multilayer structures. However, such evidence of coupling is indirect, having been obtained through techniques which can only infer the behavior of individual layers from the collective behavior of the entire multilayer structure. Conversely, the magnetizations $M$ of the individual layers in a multilayer structure can be obtained through a quantitative analysis of the structure's polarized neutron reflectivity (PNR).[14,15,16,17] In this communication we report on PNR measurements of a series of $Ga_{1-x}Mn_xAs$ based multilayer structures. Our results unambiguously show that $Ga_{1-x}Mn_xAs$ layers can couple across a non-magnetic spacer layer, and that the coupling strength is dependent on the spacer thickness.

PNR measurements were conducted using the NG-1 Reflectometer at the NIST Center for Neutron Research. A magnetic field $H$ was applied in the plane of the sample, defining a magnetic reference frame for the experiment. A neutron beam of wavelength 4.75 Å was polarized alternately spin-up (+) and spin-down (-) relative to $H$, and was incident on the sample. The non spin-flip specular reflectivities $R^{++}$ and $R^{--}$ were measured as a function of wavevector transfer $Q$.[18] Standard corrections were applied to the data to correct for background, neutron polarization efficiencies, and footprint of the incident beam. A sample's depth-dependent nuclear scattering length density $\rho(z)$, and depth-dependent magnetization component $M(z)$ parallel to $H$, can be deduced by model fitting[19] of $R^{++}(Q)$ and $R^{--}(Q)$.[14,16,17,18] In this way, we determined the thickness, Mn concentration $x$[20,21], and $M$ for the individual layers in each of the samples studied.[22]

Three 1 cm x 2 cm rectangular samples were prepared by molecular beam epitaxy on GaAs substrates[23] with the following layer structure (starting at the substrate interface): a $16^{\pm 2}$ nm bottom layer of FM $x = 0.05^{\pm 0.015}$ $Ga_{1-x}Mn_xAs$, a non-magnetic GaAs spacer, an $8^{\pm 2}$ nm top layer of FM $x = 0.05^{\pm 0.015}$ $Ga_{1-x}Mn_xAs$, and a $25^{\pm 1.5}$ nm $Al_{0.25}Be_{0.31}Ga_{0.44}As$ cap.[24] We observe that nominally[25] the three samples are structurally identical except for the spacer thickness, which is $12^{\pm 1}$, $6^{\pm 1}$, and $3^{\pm 1}$ nm, respectively.[26,27] Figure 1 shows the fitted PNR data taken out to highest $Q$, and therefore used to determine the structural parameters of the samples. These data were taken at different $H$ and $T$ (1 mT and 25 K for the 12 nm spacer sample, 1 mT and 40 K for the 6 nm spacer sample, 100 mT and 5 K for the 3 nm spacer sample) but all correspond to conditions where the magnetizations of the top and bottom $Ga_{1-x}Mn_xAs$ layers ($M_{top}$ and $M_{bot}$) are similar to each other and nonzero. For all three samples, fits to the PNR data are extremely sensitive to the presence of a spacer layer with greatly reduced $x$ (consistent with 0), and greatly reduced $M$ (also consistent with 0) in the model. Fitting error associated with these parameters suggest that the spacer has at most ~ 10 times smaller $x$, and ~ 4 times smaller $M$ than the adjacent $Ga_{1-x}Mn_xAs$ layers. Models that treat the $Ga_{1-x}Mn_xAs$ /GaAs/ $Ga_{1-x}Mn_xAs$ structure as a single layer of uniform $x$ and/or $M$ result in substantially worse fits to the data. This confirms that little or no Mn is present in the spacers of these samples.

The role of the beryllium-doped AlGaAs capping layer is to make the top and bottom $Ga_{1-x}Mn_xAs$ layers magnetically different through modulation hole-doping of the top $Ga_{1-x}Mn_xAs$ layer.[23,28] Since the FM exchange in $Ga_{1-x}Mn_xAs$ is hole mediated,[6,29] the FM transition temperature of a $Ga_{1-x}Mn_xAs$ is increased when it is adjacent to an $Al_yBe_zGa_{1-y-z}As$ layer. For our samples, this means that the top and bottom $Ga_{1-x}Mn_xAs$ layers will exhibit very different $M(T)$ curves – unless the two layers are coupled across the spacer.

In order to test for such coupling and examine how it changes with spacer thickness, we used PNR to measure $M(T)$ of the individual layers of each sample, while in a remnant magnetic state.[30] To examine the effects of anisotropy, these measurements were conducted for $H$ along each of the in-plane crystallographic directions: a magnetic hard [110], a magnetic easy [110], and the [100].[31] Because the densities of the samples do not change appreciably over the $T$ range examined, $T$-dependent changes in a sample's PNR can be solely attributed to magnetic changes in the sample. Since the magnetic properties of the sample are manifested in the differences between $R(Q)^{++}$ and $R(Q)^{--}$, it is intuitive to express the PNR data as spin asymmetry: $A(Q) = \dfrac{R^{++}(Q) - R^{--}(Q)}{R^{++}(Q) + R^{--}(Q)}$.

Model calculations show that the frequency of the $A(Q)$ oscillations is strongly dependent on the *total* magnetized thickness of the entire multilayer structure. Thus, the $A(Q)$ frequency exhibited when $M_{top}$ and $M_{bot}$ are similar is distinct from the frequency exhibited when $M_{top}$ and $M_{bot}$ are very different. The amplitude of the peaks is largely dependent on the total $M$ of the sample. Therefore, the sample's $A(Q)$ falls into one of three general categories:

1) A high frequency oscillation corresponding to low $T$, where $M_{top} \approx M_{bot}$.

2) A shorter frequency oscillation corresponding to $M_{top} \neq 0$, $M_{bot} \approx 0$.

3) $A(Q) = 0$, corresponding to $M_{top} = M_{bot} = 0$.

As an example, Figure 2 shows $A(Q)$ data at selected $T$ for each of the samples, measured with $H = 1$ mT along the hard [110] direction. The solid lines through the data points are derived from fits to $R^{++}(Q)$ and $R^{--}(Q)$. For the 12 nm spacer sample in Figure 2a, clearly different $A(Q)$ frequencies are exhibited for the two $T$, indicating that the total magnetized thickness of the sample changes between 40 and 60 K. Figure 2b indicates a similar transition between 40 and 70 K for the 6 nm spacer sample. In contrast, Figure 2c shows no change in frequency for the 3 nm spacer sample between 40 and 80 K – indicating that the total magnetized thickness does not change over this $T$ range. Figure 2 presents a compelling qualitative argument that the $M(T)_{top}$ and $M(T)_{bot}$ become more similar as the spacer layer thickness is reduced. Quantitative results obtained from fits to the PNR data are presented in Figure 3, which shows the full $M(T)_{top}$ and $M(T)_{bot}$ curves, for $H$ along each of the in-plane crystal directions. The net sample magnetizations corresponding to the curves in Figure 3 agree well with SQUID[32] magnetometry measurements of the same samples[33]. Effects of $T$-dependent anisotropy are evident for the 12 nm spacer sample and the 6 nm spacer sample, as the relationship between $M(T)_{top}$ and $M(T)_{bot}$ for the hard [110] direction is significantly different than it is for the easy [110] or the [100] direction. Such effects are not so apparent for the 3 nm spacer sample, as $M(T)_{top}$ and $M(T)_{bot}$ look comparatively similar to one another in all three directions. The differences among the samples are easily seen in Figure 4, which shows $M(T)_{bot}$ / $M(T)_{top}$ for each one. The $T$ at which this ratio goes to zero indicates the vanishing of $M_{bot}$. For all three directions, $M(T)_{bot}$ / $M(T)_{top}$ is nonzero below 50 K for the 12 nm sample, below 60 K for the 6 nm sample, and below 80 K for the 3 nm sample. Figure 4 clearly illustrates the primary finding of this paper: Regardless of magnetization direction, $M(T)_{top}$ and $M(T)_{bot}$ become progressively more similar to one another as the spacer thickness between them is reduced.

Since the three samples are virtually identical except for spacer thickness, this behavior is definitive evidence of a coupling between the $Ga_{1-x}Mn_xAs$ layers that varies in strength as the distance between them increases. Further, we conclude that the observed coupling cannot be mediated by a "magnetic short" resulting from unintended FM in the spacer layer. As discussed previously, at 100 mT and 5 K (conditions approaching magnetic saturation of $Ga_{1-x}Mn_xAs$), we observe for the 3 nm spacer sample that the spacer has greatly reduced $x$ and $M$ (i.e. consistent with zero) compared to the surrounding $Ga_{1-x}Mn_xAs$ layers. Figure 4 shows that the coupling between the $Ga_{1-x}Mn_xAs$ layers in this sample persists up to at least $T = 80$ K. Therefore, for FM bridges or shorts in the spacer to be responsible for the interlayer coupling, FM order in a layer with no observable magnetic dopant or low-$T$ $M$ would also have to persist above 80 K. Considering that $M$ for the bottom $Ga_{1-x}Mn_xAs$ layer of the 12 nm spacer sample – which has $x = 0.05$ and a sizable low-$T$ $M$ – disappears below 60 K, such a scenario is totally implausible.

We instead propose that itinerant holes are responsible for the observed spacer-dependent interlayer coupling. One possibility is that interlayer hopping of free holes results in a RKKY-like FM exchange coupling of the two $Ga_{1-x}Mn_xAs$ layers that becomes stronger as the spacer thickness is reduced. Another possibility is that the two $Ga_{1-x}Mn_xAs$ layers are electronically coupled by holes.[33] As the spacer thickness is reduced, the wavefunctions of holes in the top $Ga_{1-x}Mn_xAs$ layer may begin to significantly overlap the wavefunctions of holes in the bottom $Ga_{1-x}Mn_xAs$ layer.[34] Such an overlap could make it easier for carriers in the hole-rich top $Ga_{1-x}Mn_xAs$ layer to overcome the potential barrier of the GaAs spacer and drift into the bottom $Ga_{1-x}Mn_xAs$ layer - resulting in a more equal distribution of holes, and thereby more similar FM properties in the two $Ga_{1-x}Mn_xAs$ layers. Either of these coupling mechanisms, or a combination of the two, could explain our results.

In conclusion, we observe coupling between $Ga_{1-x}Mn_xAs$ layers across spacers as thick as 6 nm, a factor of three greater than the thickness for which it has been theoretically predicted that RKKY-like FM coupling should effectively disappear, and thicker than spacers across which interlayer coupling was experimentally inferred in References 8-13. This suggests that the carrier-mediated interaction between separated $Ga_{1-x}Mn_xAs$ layers is more robust than previously thought, a quality that may prove beneficial for device applications.

This work was supported by NSF DMR-0603752. The authors thank Chuck Majkrzak, Paul Kienzle, Shannon Watson, and Brian Maranville of NIST, and Mike Fitzsimmons of Los Alamos National Laboratory for enlightening discussions.

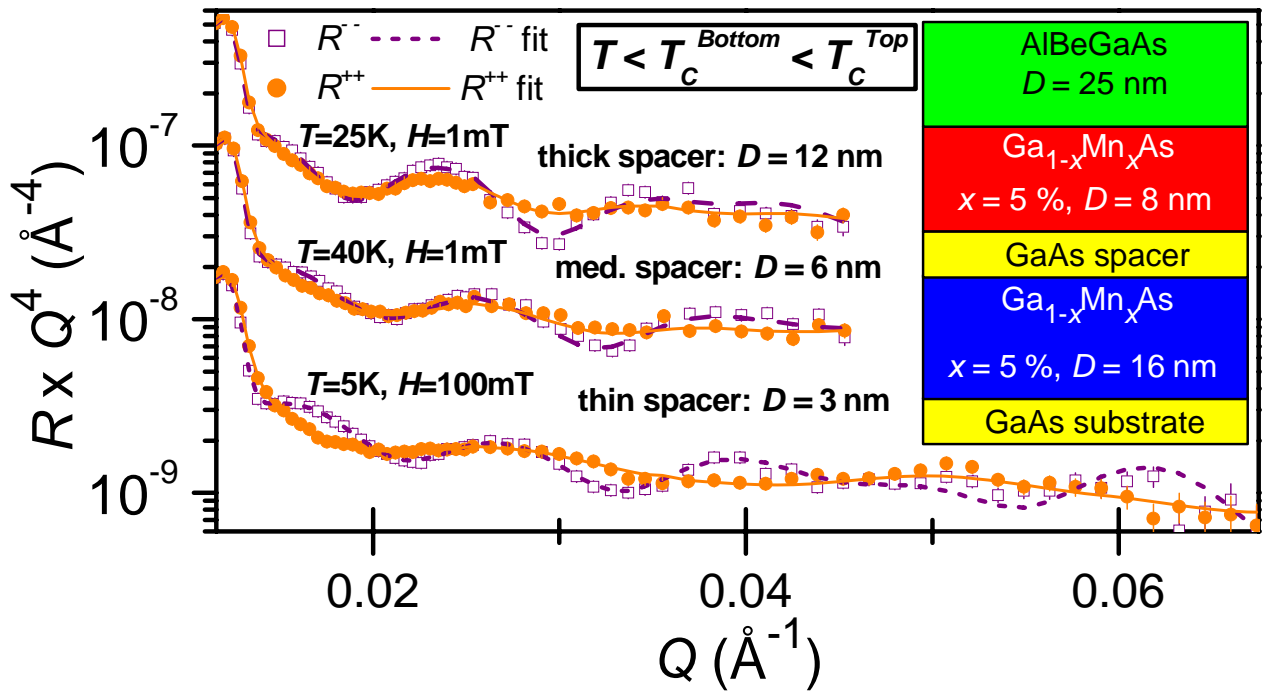

Figure 1. PNR data and fits used to determine the layer thicknesses and compositions for the samples. The data are vertically offset, and multiplied by $Q^4$ for clarity.

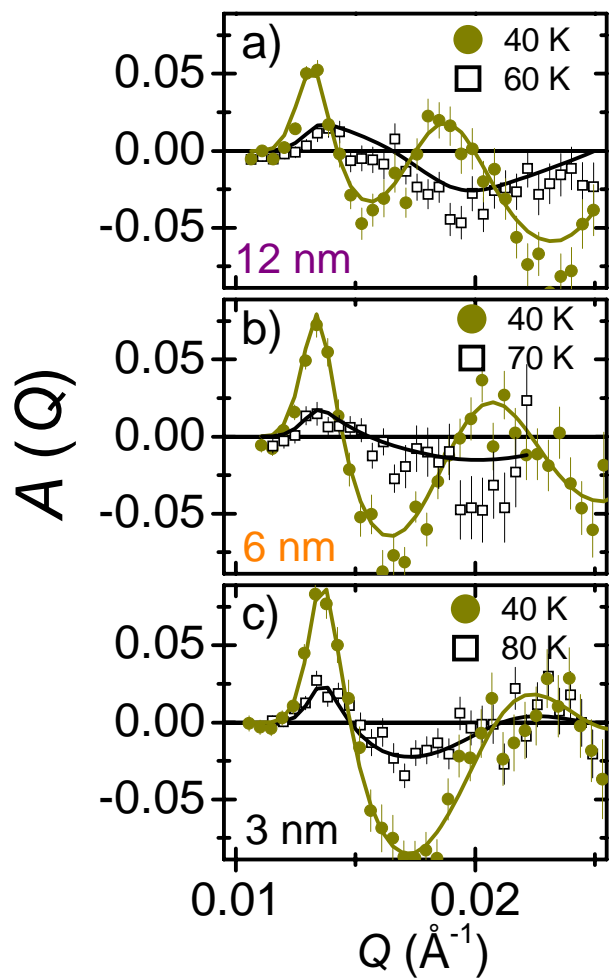

Figure 2. PNR data and fits (solid lines) expressed as spin asymmetry, for the a) 12 nm spacer sample, b) 6 nm spacer sample, and c) 3 nm sample, taken with $H = 1$ mT along the hard [110] direction.

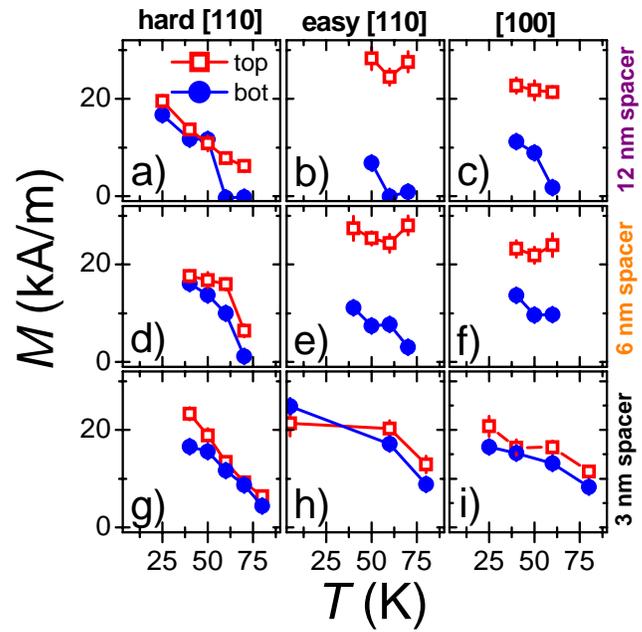

Figure 3. $M(T)$ for the individual top and bottom $Ga_{1-x}Mn_xAs$ layers for the three samples (rows), for $H$ along each of the three in-plane crystallographic directions (columns). Solid lines are guides to the eye.

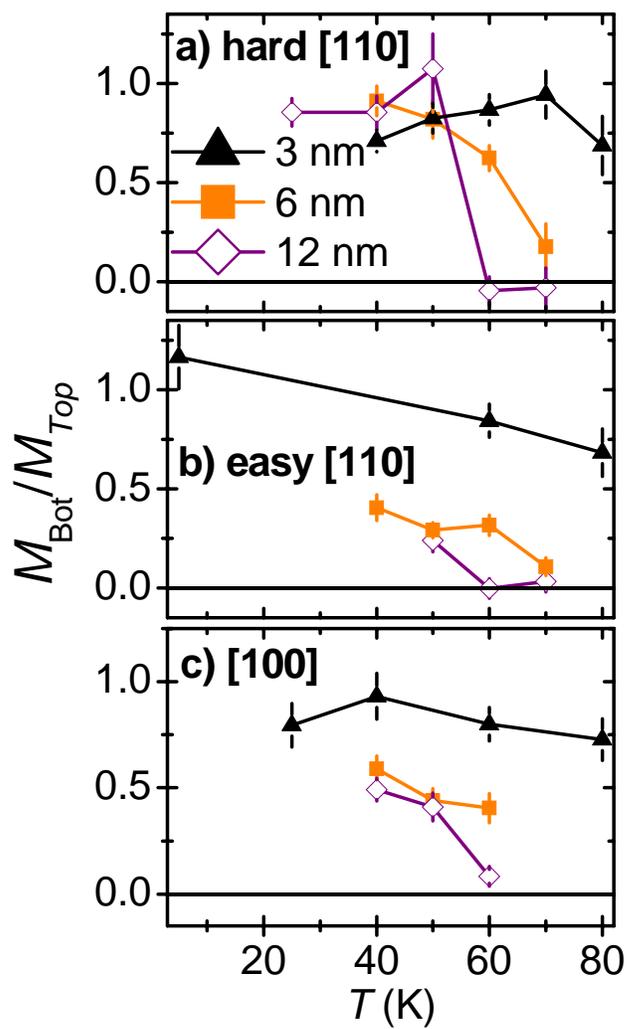

Figure 4. Ratio of the bottom and top $Ga_{1-x}Mn_xAs$ layer magnetizations for each of the samples for *H* along a) the hard [110], b) the easy [110], and c) the [100]. Solid lines are guides to the eye.